\documentstyle[12pt,psfig]{l-aa}

\begin{document}
\newcommand{\vb}{\mbox{$ \vec{\beta}$}}
\newcommand{\vbb}{\mbox{$ \vec{\beta}_b$}}
\newcommand{\gb}{\mbox{$\gamma_b$}}
\newcommand{\bb}{\mbox{$\beta_b$}}
\newcommand{\vk}{\mbox{$\vec{k}$}}
\newcommand{\dop}{\mbox{$\Gamma$}}
\newcommand{\Inuk}{\mbox{$I_{\nu}(\vec{k})\ $}}
\newcommand{\Ik}{\mbox{$I(\vec{k})$}}
\newcommand{\hn}{\mbox{$(h\nu)^{-1}$}}
\newcommand{\beq}{\begin{equation}}
\newcommand{\eeq}{\end{equation}}
\newcommand{\beqar}{\begin{eqnarray}}
\newcommand{\eeqar}{\end{eqnarray}}
\newcommand{\oJ}{\mbox{$\overline{J}$}}
\newcommand{\oH}{\mbox{$\overline{H}$}}
\newcommand{\oT}{\mbox{$\overline{T}(\mu)$}}
\newcommand{\oeta}{\mbox{$\overline{\eta}$}}
\newcommand{\feta}{\mbox{$F_{\eta\chi}(\mu)$}}
\thesaurus{11(02.18.8;02.19.2;03.13.4;11.01.2;11.19.1)}
\title{Anisotropic illumination of AGN's accretion disk by a non thermal
  source : I General theory and application to the Newtonian geometry}
\author{G. Henri \and P.O. Petrucci}
\institute{Laboratoire d'Astrophysique, Observatoire de Grenoble,B.P 53X,
F38041 Grenoble Cedex, France}
\date{Received ??; accepted ??}
\maketitle
\markboth{G.Henri \& P.O. Petrucci: Anisotropic illumination of AGN's I}{}

\begin{abstract}
We present a model of accretion disk where the disk luminosity is 
entirely due to the reprocessing of hard radiation impinging on the disk. 
The hard radiation itself is emitted by a hot point source above the 
disk, that could be physically realized by a strong shock terminating 
an aborted jet. This hot source contains ultrarelativistic leptons 
scattering the disk
soft photons by Inverse Compton (IC) process. Using a simple formula to 
describe the IC process in an anisotropic photon field, we derive a 
self-consistent angular distribution of soft and hard radiation
 in the Newtonian geometry. The radial profile of the disk effective 
temperature is also univoquely 
determined. The high energy spectrum can be calculated for a given lepton
distribution. This offers an alternative picture to the standard accretion 
disk emission law. We discuss the application of this model to Active 
Galactic Nuclei, either for reproducing individual spectra, or for 
predicting new scaling laws that fit better the observed statistical 
properties.   
\end{abstract}

\section{Introduction}

With the development of high energy telescopes, it has been recognized 
that Active Galactic Nuclei are the most powerful emitters of high 
energy radiation in the Universe. However, the detailed production 
mechanism is still a matter of debate. For radio loud AGN, 
the detection of very high 
energy radiation, in the GeV (von Montigny et al. \cite{vMont95}) and even TeV (Punch et al. 
\cite{Punch92}; Quinn et al. \cite{Quin96}) ranges, proves 
the existence of ultrarelativistic
 particles, probably associated with a relativistic jet (e.g. 
 Begelman et al. \cite{Begel84}; Dermer \& Schlickeiser \cite{DermS92}). No 
such conclusion can be drawn up to now for radio-quiet objects such as 
Seyfert galaxies, since their high energy spectrum is apparently cut-off 
above a few hundred keV (Jourdain et al. \cite{Jou92a}; Maisack et 
al. \cite{Mais93}; Dermer \& Gehrels \cite{DermGehr95}). Although this
radiation could be produced  
by direct synchrotron mechanism, it is more often assumed that it comes 
from the Comptonization of soft photons by high energy electrons or 
pairs. Two
classes of models have been proposed so far: Comptonization by a thermal,
mildly relativistic, plasma, resulting in a lot of scattering events
associated with small energy changes, or Inverse Compton (IC) process
by one or few scattering events from a highly relativistic, non 
thermal particles distribution, which can result from a pair cascade.\\
\indent
Detailed observations in the X-ray range by the Ginga
satellite have shown that a simple power law is unable to fit the X-ray 
spectrum of Seyfert galaxies. Rather, the spectra are better reproduced 
by a complex
superposition of a primary power law, with an index $\alpha \simeq 
0.9-1.0$, a reflected component from a cold
thick gas, a fluorescent Fe K$\alpha$ line and an absorption edge by a warm 
absorber (Pounds et al. \cite{Poun90}; Nandra \& Pounds \cite{Nand94}).
The second and third components could be produced by the reflection of
primary hard radiation on an accretion disk surrounding the putative
massive black hole powering the AGN (Lightman \& White \cite{LW88}; 
George \& Fabian \cite{GF91}; Matt, Perola \& Piro \cite{MattP91}). 
This has led to consider 
various geometries where the hot source is located above the disk and 
reilluminates it, producing the observed reflection features. The hot 
source can be a non-thermal plasma (Zdziarski et al. \cite{Zdz90}), or 
a thermal hot corona covering the disk (Haardt \& Maraschi \cite{HaaMar91},
\cite{HaaMar93}; Field \& Rogers \cite{Fiel93}). \\ 
\indent
In another context, some observational facts have motivated the 
 development of so-called reillumination models, where high energy 
radiation 
reflected on a cold surface (presumably again the surface of an accretion 
disk), produces a fair fraction of thermal 
UV-optical radiation. Firstly, long term observations have shown that for 
some Seyfert galaxies, such as NGC 4151 (Perola et al. \cite{Per86}) 
and NGC 5548 (Clavel et al. \cite{Clav92}), UV and optical 
luminosities were varying simultaneously, and correlated with X-ray 
variability on time scales of months, whereas the rapid, short-scale X-ray 
variability was not seen in optical-UV range. This is in contradiction 
with the predictions of a standard, Shakura-Sunyaev (SS) accretion disk 
model (Shakura \& Sunayev \cite{Shak73}), 
where any perturbation causing optical variability should cross the disk 
at most at the sound velocity, producing a much larger lag between 
optical and UV than what is actually observed. Rather, these observations 
support the idea that optical-UV radiation is largely produced by 
reprocessing of X-rays emitted by a small hot source, the UV and 
optical radiation being emitted at larger distances. The main problem 
is that the apparent X-ray luminosity is usually much lower than 
the optical-UV continuum contained in the Blue Bump, whereas one would 
expect about the same intensity in both components if half of the primary 
hard radiation is emitted directly towards the observer and the other 
half is reprocessed by the disk.\\
\indent
  In many cases also, the
equivalent width of the Fe K$\alpha$ line requires more impinging
radiation than 
what is actually observed if explained by the reflection model
 (Weaver et al. \cite{Weav95}, Nandra et al. \cite{Nand97}).
 As an explanation, Ghisellini
et al. (\cite{Ghis91}), hereafter G91,  have proposed that the anisotropy 
of soft radiation could lead to
an anisotropic IC emission, with much more radiation being scattered
backward than forward. Due to the complexity of their calculation, they
have restricted themselves to the emission by a hemispheric bowl 
(equivalent to an infinite plane), that 
could model a flared accretion disk with a constant temperature. The 
thermal disk-corona model faces the same kind of difficulties, for it 
predicts nearly the same luminosity in X-ray and UV ranges. A
possible solution could imply a patchy corona, a fair part of the UV
luminosity being emitted by internal dissipation in the disk (Haardt \&
al. \cite{HMG95}). 
In the cases where X and UV luminosities are comparable however, it is 
difficult to explain very rapid X-ray variability as the corona must cover 
a large part of the disk. \\
\indent
  Although the 
spectral break observed by OSSE around 100 keV seems to favor 
thermal models and disprove the simplest pair cascade models, such a 
break could also be obtained by a relativistic particles distribution with 
an appropriate upper energy cut-off, such can be provided for example 
by pair reacceleration to avoid pair run-away (Done et al. 
\cite{Done90}, Henri \& Pelletier \cite{HP91}). The aim of
this paper is to reconsider the reillumination by a non thermal,
optically thin IC 
source, taking properly into account the disk geometry and the 
anisotropic 
distribution of photons. We first establish a simple expression to
evaluate the power emitted by a single particle scattering photons by 
Compton mechanism in the Thomson regime in an arbitrary soft
photon field. The formulae require only the computation of the 
components of 
the relativistic radiation tensor, or equivalently 
the Eddington parameters for
an axisymmetric field. We then develop  
a self consistent model where the
emission of the disk is entirely due to the reprocessing of hard
radiation, produced itself by IC process in a hot point source located 
above the disk. In this case a unique angular distribution of hard radiation 
and a unique (properly scaled) disk
temperature radial profile are predicted.  We discuss then the possible
physical mechanisms for such a situation and its implication for the
overall characteristics of AGNs, both for individual spectra and for 
statistical properties. We derive new 
scaling laws for luminosity and central temperature as a function of the 
mass. We show that the predictions of the model are sensitively 
different from 
the standard ones, and that they could better explain the observations. 

\section{Anisotropic Inverse Compton process}
\subsection{Total power emitted by a single particle}

We first establish useful formulae to compute the Inverse Compton (IC) 
emissivity of a particle in an arbitrary photon field, in the Thomson 
regime.
We consider the case of a relativistic charged particle with mass $m$,
velocity $\vec{v}=\beta c \vk_0$, and Lorentz factor $\gamma =
(1-\beta ^2)^{-1/2}$, in a soft photon field characterized by the 
specific intensity distribution \Inuk. $\vk$ and $\vk_0$ are 
respectively the unit 
vectors along the photon and the particle velocity. 
We assume that the Thomson
approximation is valid, that is $\epsilon \gamma \ll 1$ where $\epsilon =
{h\nu / mc^2}$ is the soft photon energy in units $mc^2$. In this
limit, the rate of energy transferred from the particle to the photons is: 
\beq
\frac{dE}{dt} = P_+ - P_- \label{dEdt}
\eeq
where
\beq
P_+ = \sigma_T \int \Inuk(1-\beta \vk_0 . \vk) d\Omega d\nu   \label{P+}
\eeq
and
\beq
P_- = \sigma_T \gamma^2 \int \Inuk(1-\beta \vk_0 . \vk)^2 d\Omega d\nu 
\label{P-}
\eeq
are respectively the power brought by the incident photons and carried
out by the scattered ones. Here $\sigma_T = 6.65\,10^{-25} \mbox{cm}^{2}$ is 
the usual Thomson cross section.\\
\indent
To transform these expressions, it is useful to consider the
decomposition of the intensity field $\Ik = \displaystyle\int \Inuk d\nu$ 
over the spherical harmonics basis:
\beq
\Ik = \int \Inuk d\nu = \sum_{\stackrel{m=-l}{l=0}}^
{\stackrel{l=\infty}{m=l}} c_{lm} Y_l^m(\vk)  \label{Inuk}
\eeq
where, due to the orthonormality condition
\beq
\int Y_l^m(\vk)Y_{l'}^{m'*}(\vk) d\Omega = 
\delta_{ll'}\delta_{mm'}  \label{ortho},
\eeq
the coefficients $c_{lm}$ are given by:
\beq
c_{lm} = \int \int \Inuk Y_l^{m*}(\vk) d\Omega d\nu   \label{clm}
\eeq
(here * denotes the complex conjugate and $\delta_{ll'}$ the usual
Kronecker symbol equal to 1 if $l = l'$ and 0 or else). Note that because
\Inuk is  
real, one has the conjugation relationship $c_{l-m} = c_{lm}^*$.\\ 
Now one can write $\vk_0 . \vk = \cos \alpha$, where $\alpha$ is
the angle between the particle velocity and the incident photon, and use
the following expansion formulae:
\beqar
\cos \alpha & = & \frac{4 \pi}{3}\sum_{m = -1}^{m = +1} Y_1^m(\vk_0) 
Y_1^{m*}(\vk) \label{cosalpha} \\ 
\cos^2 \alpha & = & \frac{8 \pi}{15}\sum_{m = -2}^{m = +2} Y_2^m(\vk_0)
Y_2^{m*}(\vk) + \frac{4\pi}{3} Y_0^0(\vk_0)Y_0^{0*}(\vk). \label{cos2alpha}
\eeqar
Inserting Eq. (\ref{Inuk}), (\ref{cosalpha}) and (\ref{cos2alpha}) in Eq. 
(\ref{dEdt})-(\ref{P-}), and using the relation (\ref{ortho}),
 one gets finally:
\beqar
P_+ & = & 4\pi \sigma_T[c_{00}Y_0^0-\frac{\beta}{3}\sum_{m = -1}^
{m = +1} c_{1m}Y_1^m] \label{P+1}, \\
P_- & = & 4\pi \sigma_T\gamma^2[c_{00}(1+\frac{\beta^2}{3})Y_0^0
\nonumber \\
 & & - \frac{2\beta}{3}\sum_{m = -1}^{m = +1}c_{1m}Y_1^m+ 
\frac{2\beta^2}{15}\sum_{m = -2}^{m = +2}
c_{2m}Y_2^m], \label{P-1} \\
\frac{dE}{dt} & = & - 4\pi \sigma_T\gamma^2\beta[c_{00}
\frac{4\beta} {3}Y_0^0 \nonumber \\
 & & - \frac{1+\beta^2}{3}\sum_{m = -1}^{m = +1}
c_{1m}Y_1^m + \frac{2\beta}{15}\sum_{m = -2}^{m = +2} 
c_{2m}Y_2^m]. \label{dEdt1}
\eeqar
Thus the computation of the power emitted in any direction requires the
computation of the 9 components $c_{lm}$ $\{ l = 0,1,2; -l\le m \le +l\}$ 
of the radiation field (related 
to the 9 independent components of the relativistic radiation tensor).
These formula can further be simplified in the important case of an
axisymmetric field. There $c_{lm} = c_l \delta_{0m}$, and the relevant
spherical harmonic functions are given by:
\beqar
Y_0^0(\vk_0) & = & {1 \over \sqrt{4 \pi}} \\
Y_1^0(\vk_0) & = & {3 \over \sqrt{4 \pi}} \mu \\
Y_2^0(\vk_0) & = & {5 \over \sqrt{16 \pi}} (3\mu^2 -1),
\eeqar
where $\mu = \cos\theta_0=\vk_0 .\vec{z}$, $\vec{z}$ being the unit vector
of the vertical axis. Using the Eddington parameters
\beqar
J & = & {1 \over 2}\int \Inuk d\mu d\nu = \frac{1}{\sqrt{4 \pi}}  c_{00}
\nonumber \\
H & = & {1 \over 2}\int \Inuk \mu d\mu d\nu = \frac{1}{\sqrt{12 \pi}}  
c_{10} \\
K & = & {1 \over 2}\int \Inuk \mu^2 d\mu d\nu = \frac{1}{36 \pi} (c_{00}+
\frac{2}{\sqrt{5}} c_{20}), \nonumber
\eeqar
one gets finally:
\beqar
P_+ & = & 4 \pi \sigma_T[J-\beta H \mu]\\
P_- & = & 2 \pi \sigma_T \gamma^2 [2 J+\beta^2 (J-K) \nonumber \\
 & & -4\beta H \mu +\beta^2 (3K-J)\mu^2]\\
\frac{dE}{dt} & = & - 2 \pi \sigma_T \gamma^2 \beta[\beta (3J-K)
\nonumber \\
 & & -2(1+\beta^2)H \mu +\beta(3K-J)\mu^2]    
\eeqar
These expressions appear like simple polynomials of order 2 in $\mu$,
involving only the calculation of the three Eddington parameters.
In the case of a ultrarelativistic particle $\gamma \gg 1$, they take the
form:
\beqar
P_+ & = & 4 \pi \sigma_T J (1-\eta \mu)           \label{pplus} \\
P_- \simeq -\frac{dE}{dt}& = & 2 \pi \sigma_T \gamma^2 J \feta \label{pmin}
\eeqar
where we introduce the following notations:
\beqar
  \eta & = & \frac{H}{J} \nonumber \\
  \chi & = & \frac{K}{J} \\
  \feta & = & [(3-\chi) -4 \eta \mu +(3\chi - 1)\mu^2].
\eeqar

\subsection{ Emitted spectrum}
\indent
Although the total emitted power can be cast into the above relatively
simple forms, there is no such simplification for the spectrum of the
emitted radiation. This is because photons with a given energy can be 
produced by different combinations of initial energy, incident angle, and 
scattering angles and thus the exact spectrum depends on the detailed 
form of the soft photon distribution and not only on the Eddington 
moments. A complete calculation requires the integration of the 
Klein-Nishina cross-section over photon energies, particle energies 
and relative angles.
 However, in the case of IC scattering of a single particle on a monoenergetic,
  isotropic soft photon distribution, a convenient approximation is often to take 
 a $\delta$-function
 \beq
 {\stackrel{.}{n}}_s\delta(\epsilon'-\langle\epsilon'\rangle) \label{dirac}
 \eeq
where $\langle\epsilon'\rangle = \displaystyle\frac{4}{3} 
\gamma^2\epsilon_s$ is the mean energy of 
Comptonized photons.\\
In the case of a relativistic distribution, this approximation is
reasonable if the width of soft photon energy spectrum is much less than
the width of the particle energy distribution. As we shall see, the soft photon
spectrum predicted by the present model is close to a blackbody, and we
will keep this kind of approximation. One can easily generalize
expression (\ref{dirac}) to an 
arbitrary soft photon field by taking the appropriate expression for 
the Comptonized photons mean 
energy. It is obtained by dividing the emitted power (Eq. (\ref{P-1}))
 by the rate of 
photon scattering.
The latter is given by
\beq
{\stackrel{.}{n}}_s = \sigma_T \int \Inuk \hn (1-\vb . \vk) d\Omega d\nu
\eeq
A calculation quite similar to that of the previous paragraph gives:
\beq
{\stackrel{.}{n}}_s  = 4\pi \sigma_T[d_{00}Y_0^0(\vk_0)-
\frac{\beta}{3}\sum_{m = -1}^{m = +1} d_{1m}Y_1^m(\vk_0)]
\eeq
where $d_{lm} =\displaystyle \int \int \Inuk \hn Y_l^{m*}(\vk) d\Omega d\nu$ is
calculated with the photon number flux instead of the energy flux.\\ 
In the case of
an axisymmetric photon field again, one can simplify this expression
using the photon number Eddington parameters:
\beqar
\oJ & = & {1 \over 2}\int \Inuk \hn d\mu d\nu = (4 \pi)^{-1/2} d_{00}\\
\oH & = & {1 \over 2}\int \Inuk \hn \mu d\mu d\nu = 
(12 \pi)^{-1/2} d_{10}\\
\oeta & = & \frac{\oH}{\oJ},
\eeqar
to get:
\beq
{\stackrel{.}{n}}_s = 4\pi \sigma_T(\oJ - \beta \oH \mu).
\eeq
The mean photon energy of the emitted radiation is thus:
\beqar
\langle\epsilon'\rangle & = & \frac{P_-}{{\stackrel{.}{n}}_s} \nonumber\\
                        & = & \gamma^2 \frac{[2J+\beta^2(J-K)
-4\beta H \mu +\beta^2 (3K-J)\mu^2]}{2(\oJ-\beta \oH \mu)}.
\eeqar
For ultrarelativistic particles, these expressions become
\beqar
{\stackrel{.}{n}}_s & = & 4\pi \sigma_T \oJ (1 - \oeta \mu) \label{dndt} \\
\langle\epsilon'\rangle
& = & \gamma^2 \langle\epsilon_s\rangle A(\mu) \label{meaneps}
\eeqar
 where $\displaystyle \langle\epsilon_{s}\rangle = \frac{J}{\overline{J}}$ 
 is the mean energy of incident soft photons and 
 \beq
 A(\mu) = \frac{\feta}{2(1-\overline{\eta}\mu)} \label{amu}
 \eeq
  is an 
 angle-dependent numerical factor. For an isotropic photon distribution, 
 $\eta = \oeta = 0$ and $\chi = 1/3$ and
 one gets the familiar result $A(\mu) = 4/3$.
Just as in the isotropic case, one can approximate the spectrum by a
Dirac distribution of Eq. (\ref{dirac}), if most of the emitted energy comes 
from a 
restricted range of soft photons energy and direction. One can expect 
this to be a good approximation if the particle energy distribution is 
broad enough, so that the intrinsic broadening due to photon energy 
distribution is negligible, except near the spectrum energy cut-offs. 

\subsection{Emission by a relativistic particles distribution}
\indent
The previous formulae can be applied to the case of a
relativistic particles distribution. For sake of
simplicity, we will restrict ourselves to the case of an axisymmetric
distribution $f(\gamma,\mu)$, which represents the particle number
(integrated over the volume) per energy and angle cosine interval.
 Axisymmetry is automatically insured at first
approximation by the cyclotron precession around a small magnetic field 
aligned with the symmetry axis of the radiation field.
 For an isotropic distribution, $f(\gamma,\mu)=n(\gamma)/2$,
where $n(\gamma)$ is the particle energy distribution. The plasma is
assumed to be optically thin, such that every particle experiences the
same radiation field. This point will be further discussed in Section
(\ref{Discussion}). 
 We assume further that the low
energy cut-off is high enough to make Eq. (\ref{pplus}) - (\ref{pmin})
 valid, whereas the
high-energy cut-off is still in the Thomson regime.

\subsubsection{The integrated power}
 The integrated
plasma emissivity can be written
\beq
\frac{dP}{d\Omega} = (2\pi)^{-1}\int f(\gamma,\mu) \frac{dE}{dt} d\gamma.
\eeq
Inserting Eq. (\ref{pmin}) yields
\beq
\frac{dP}{d\Omega} = \sigma_T J \int_{\gamma_{min}}^{\gamma_{max}}\feta
f(\gamma,\mu) \gamma^2 d\gamma.
\eeq
 Defining the normalized angular distribution function
\beq
        g(\mu) =\frac{1}{N \langle \gamma^2 \rangle}
        \int_{\gamma_{min}}^{\gamma_{max}} f(\gamma,\mu)   
        \gamma^2 d\gamma  
\eeq
where $N$ is the total relativistic particle number and
\beq
\langle\gamma^2\rangle = \frac{1}{N}\int_{-1}^{+1}d\mu 
\int_{\gamma_{min}}^{\gamma_{max}}
f(\gamma,\mu) \gamma^2 d\gamma 
\eeq
is the mean quadratic Lorentz factor, one can rewrite this expression 
under the form:
\beq
\frac{dP}{d\Omega} = 
\sigma_T N \langle\gamma^2\rangle J g(\mu)\feta.
\label{dpdo} 
\eeq
The anisotropy of emitted radiation appears thus simply as the product of
an anisotropy factor of the particles distribution $g(\mu)$ times the
anisotropy factor \feta  of the radiation field.  
For an isotropic particles distribution, $g(\mu) = 1/2$.
 An interesting case is that of a plasma moving relativistically with a
bulk velocity \vbb 
 and a corresponding Lorentz factor \gb, such can exist in 
superluminal radio-sources (Marcowith et al. \cite{Marc95}). The particle Lorentz factors in the
observer frame $\gamma$ and in the plasma rest frame $\gamma'$ are linked 
by the relation:
\beq
\gamma =\dop \gamma'
\eeq
where $\dop = [\gb(1-\bb\mu)]^{-1}$ is the usual Doppler factor. Using the
Lorentz invariance of $f(\gamma,\mu)/\gamma^2$, one
gets :
\beq
g(\mu) = \frac{\dop^5}{\gb^3(1+\bb^2)}.
\eeq
Alternatively, one can express the results in terms of mean quadratic
Lorentz factor in the plasma frame 
$\langle\gamma^{'2}\rangle ~=~\langle\gamma^2\rangle {\gb^3(1+\bb^2)}$
 to get:
 \beq
\frac{dP}{d\Omega} = \sigma_T N \langle\gamma^{'2}\rangle 
\dop^5 \feta.
\eeq
It should be stressed that the above formula holds for the emissivity at 
a given point at rest with respect to the observer, such as a small 
volume of a continuous jet. If one follows the plasma in its motion, as 
in the case of a traveling blob, the reception time interval $dt_R$ is 
related to the rest time $dt$ by $dt = \dop dt_R$ and an extra Doppler 
factor $\dop$ must be added. In any direction, the emitted power will 
vary as $\dop^{6}$.

\subsubsection{Emitted spectrum}
One can also derive the emitted spectrum in a given direction by 
integrating the single particle emissivity over the particle 
distribution $f(\gamma,\mu)$, that is
\beq
 \frac{dP}{d\Omega d\epsilon'} = 
\frac{1}{2\pi}\int_{\gamma_{min}}^{\gamma_{max}}
 \epsilon' \frac{dn}{dt d\epsilon'}f(\gamma,\mu) d\gamma
\eeq
Using the $\delta$ approximation of Eq. (\ref{dirac}) to integrate over 
$\gamma$, and the expressions (\ref{dndt}) - (\ref{meaneps}) , one gets:
\beqar
  \frac{dP}{d\Omega d\epsilon'} & = &\sigma_T (\oJ-\oH
  \mu)\left(\langle\epsilon_s\rangle A(\mu)\right)^{-1/2} 
  \epsilon'^{1/2}\times \nonumber \\
  & &  f\left(\gamma =\left(\frac{\epsilon'}{\langle \epsilon_s \rangle 
  A(\mu)}\right)^{1/2}\right) \label{spectrehe}
\eeqar

\subsection{Application to a conical photon field}
\label{paraappli}
\indent
In this section, we illustrate the simplification brought by the 
present formalism, by redressing the problem of IC process in a semi- 
isotropic photon field. This question has already been treated 
in G91, but the use of non analytical integrals 
requires numerical computation. Here we show that 
their results can be very easily recovered in the present formalism, 
leading to simple analytical expressions.

We consider in fact the more general case of a radiation emitted in a 
cone with opening 
angle $\theta$, with constant specific intensity inside the cone and 
a null intensity outside. This would be 
relevant for example for an isothermal disk with a finite radius, or 
at the vicinity of a star without limb darkening. The case $\theta = 
\pi/2$ corresponds of course to the case studied in G91.
In this case, elementary integration gives the following Eddington parameters
\beqar
J & = & \frac{cU_{iso}}{8\pi} (1-\cos \theta) \nonumber \\
H & = & \frac{J}{2}(1 + \cos \theta)\\
K & = & \frac{J}{3}(1 + \cos \theta + \cos^2 \theta)\nonumber
\eeqar
 where $U_{iso}$ denotes the
energy density emitted by the full sphere as in G91. 
Equations (\ref{pplus}), (\ref{pmin}) read now:
\beqar
P_+ & = & \sigma_T c \frac{U_{iso}}{2} (1-\cos \theta)[1-\frac{\beta}{2} 
\mu]\\
P_- & = & \sigma_T c \frac{U_{iso}}{4}\gamma^2  (1-\cos \theta)  
[2+\frac{\beta^2}{3}(2 + \cos \theta + \cos^{2}\theta)\nonumber \\
 & & - 2 \beta  \mu
(1 + \cos \theta) + \beta^{2} \mu^{2} \cos \theta (1+ \cos\theta)]\\
\frac{dE}{dt} & = & - \sigma_T c \frac{U_{iso}}{4}\gamma^2
\beta [\frac{\beta}{3}(8 - \cos \theta - \cos^{2} \theta) \nonumber\\
& & - \mu (1 + \cos \theta)(1+\beta^2) + \beta \mu^{2} \cos \theta
(1 + \cos \theta)]    .
\eeqar
The integrals in Eq. (5) of G91, corresponding to $\cos \theta = 0$, appear
 thus like simple cosine functions. The 
ratio between the (minimal) power emitted in the forward direction and 
the (maximal) power emitted in the backward direction is:
\beq
\frac{P_{-}(\mu = -1)}{P_{-}(\mu = +1)} = \frac{3+3 \beta 
(1 + \cos \theta)
+ \beta^{2} (1 + \cos \theta + \cos^{2}\theta)}{3 - 3 \beta 
(1 + \cos \theta)
+ \beta^{2} (1 + \cos \theta + \cos^{2}\theta)}
\eeq  
In the limit $\beta \simeq 1, \theta = \pi/2$, one gets a factor 7 as 
found numerically by G91.
One can also easily evaluate the ratio $R$ between
the total power emitted in the lower hemisphere and the upper one by 
an isotropic monoenergetic particles distribution:
\beq
        R \equiv \frac{\int_{-1}^0 P_-d\mu}{\int_0^{+1} P_-d\mu}
=\frac{(2\beta^2+3\beta (1+\cos \theta)+6)}{(2\beta^2-3\beta 
(1+\cos \theta) +6)},
\eeq
with $\displaystyle R \rightarrow \frac {11+3\cos \theta}
{5 - 3\cos \theta}$
 when $\beta \rightarrow 1$. Once again, one can recover G91's result 
 by setting $\cos \theta = 0$, giving $ R = 2.2$. The maximally 
anisotropic case corresponds to $\cos \theta = 1$ (point source) and 
gives $R = 7$. It is thus the absolute maximal ratio between the power 
radiated in two hemispheres by an isotropic particles distribution in any
 photon field. 

\section{The self-consistent reilluminated disk:the Newtonian case}
\subsection{Assumptions of the model}
\begin{figure*}
  \psfig{width=17cm,height=4cm,angle=-90,file= 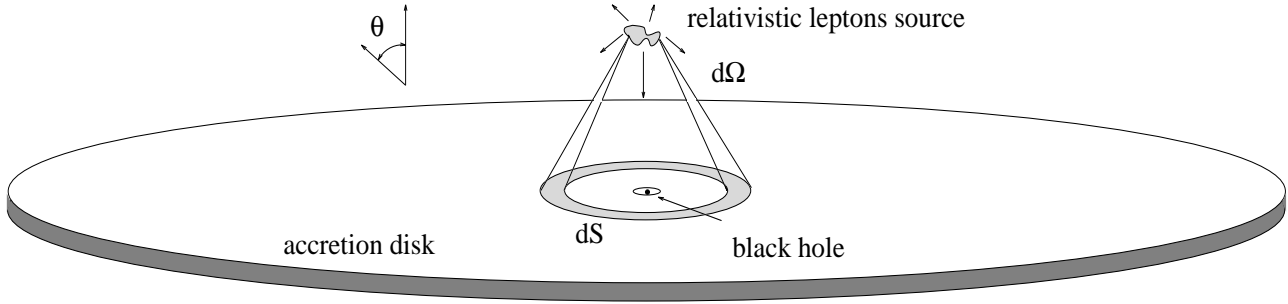}
\caption[]{The general pictures of the model. We have drawn straight
  trajectory of a beam of photons, emitted by the hot source in a solid
  angle $d\Omega$, and absorbed by a surface ring of the disk $dS$. We
  use the notation $\mu=\cos\theta_0$
           }\label{image}  
 \end{figure*}
 We consider now a self-consistent model where Inverse Compton process 
 takes place on soft photons from the accretion disk, which are 
 themselves emitted as thermal radiation due to the heating of the disk 
 by hard radiation. The disk is modelized by an infinite slab radiating
 isotropically like a black-body at the same equilibrium temperature 
(Fig. \ref{image}). The high 
 energy source is assumed to be an optically thin plasma of highly 
 relativistic leptons, at rest at a given distance $Z_0$ above the disk 
axis. 
 Its size is small enough to be considered as a point source. 
We consider a Euclidean geometry so that 
 there is no curvature of photon geodesics. The general relativistic case 
 will be studied in an accompanying paper (Petrucci \& Henri 1997). 
 The particle distribution is assumed to be isotropic. 
 As long as the disk emission is concerned, 
 there is no need to specify the energy distribution
 giving rise to IC process. However, the spectrum of 
 high energy radiation will be determined by this distribution. This 
 will be developed in section \ref{AGN}.
 
 \subsection{The self consistent solution}
 We shall see now that the above problem admits in fact a unique 
 self-consistent solution in 
 conveniently scaled variables. For a given disk emissivity, the power 
 emitted per unit solid angle by 
 the high energy source is given by Eq. (\ref{dpdo}), where $g(\mu) = 
1/2$  and the Eddington 
 parameters are to be calculated with the disk emissivity. Integrating over solid 
 angle, one gets the total high energy luminosity:
 \beq
        L_t = \frac{16 \pi}{3}\sigma_T N \langle\gamma^2\rangle J. 
        \label{Lt}
 \eeq
 Thus, the power per unit angle can be expressed as:
 \beq
\frac{dP}{d\Omega} = \frac{3L_t}{32 \pi}\feta.
\label{dpdo2}
\eeq

  The Eddington coefficients can be at turn
 calculated if one knows the disk emissivity. Under the hypothesis 
that the disk reprocesses the whole 
 radiation impinging on it, it is determined by equating the power 
absorbed and emitted by a 
 surface element $dS$ of the disk at a radius $r$ (cf. Fig. (\ref{image})):
 \begin{equation}
        F(r) dS = \frac {dP}{d\Omega}d\Omega =\mu'^3 
          \frac {dP}{d\Omega}\frac 
        {dS}{Z_0^2} \label{frds} 
 \end{equation}
 where $\mu'=-\mu = Z_0(r^2+Z_0^2)^{-1/2}$ is the cosine of the impinging 
 angle of radiation ($\mu'$ varies from 1 to 0).\\
 From equation (\ref{dpdo2}) and (\ref{frds}), one gets:
 \begin{equation}
        F(r) = \frac{3L_t}{32\pi Z_0^2}F_{\eta\chi}(-\mu')
        \mu'^3
        \label{flux}
 \end{equation}
 Under the assumption of isotropic emissivity of the disk, one gets the 
 specific intensity due to the reemission by the disk at the radius $r$ 
toward the source: 
 \begin{equation}
 I(\mu)=\frac{F(r)}{\pi}=\xi\mu^3F_{\eta\chi}(-\mu)
 \end{equation}
 where we define the dimensionless parameter 
\beq
\xi \equiv \frac{3 L_t}{32\pi^2Z_0^2J}. \label{ksi}
\eeq
  Inserting this expression in the definition of 
 Eddington parameters, one gets the following linear system
 \begin{eqnarray}
       1& = & \xi (G_0 (1-\chi) + 4 G_1 \eta + G_2 (3\chi-1))\nonumber\\
       \eta& = & \xi (G_1 (1-\chi) + 4 G_2 \eta + G_3 (3\chi-1)) \\
       \chi& = & \xi (G_2 (1-\chi) + 4 G_3 \eta + G_4 (3\chi-1))\nonumber
 \end{eqnarray}
 where 
 \beq
    G_n = \frac{1}{2} \int_{\mu_{min}}^{\mu_{max}}\mu^{n+3} d\mu 
    = \frac{1}{2(n+4)}(\mu_{max}^{n+4}-\mu_{min}^{n+4}).
 \eeq
 This homogeneous system admits a non trivial solution only if the 
 determinant is set to zero, which 
 gives a cubic equation in $\xi$. One can then determine the angular 
 parameters $\eta$ and $\chi$, and the emitted flux with
 Eq. (\ref{flux}). In the case of an infinite slab, one gets simply $G_n
 =\displaystyle 
 \frac{1}{2(n+4)}$. The numerical solutions of the cubic equation are then  
$\xi =$ 
 1.449, 48.136 and 7786.45. Only the first one is compatible with the 
 physical constraints $\eta \le 1$, $\chi \le 1$. The solutions 
 of the system are approximately 
 $\eta = 0.82288$ and $\chi = 0.69957$.\\
 With these values, the high energy emissivity has the following 
universal angle dependence:
 \begin{equation}
        \frac{dP}{d\Omega}(\mu) = L_t(0.0686-0.0982\mu+0.0328\mu^2),
 \end{equation}
 and thus the illuminated disk has the corresponding emissivity law:
 \beqar
 F(r) & = & \sigma T_{eff}^4(r) \nonumber \\ 
      & = &\frac{L_t Z_0}{(Z_0^2+r^2)^{3/2}}\times \nonumber \\ 
    & &
    \left(0.0686+\frac{0.0982}{\left(1+\left(\frac{r}{Z_0}\right)^2
    \right)^{1/2}}+\frac{0.0328}{1+\left(\frac{r}{Z_0}\right)^2}\right).  
 \label{emiss}
 \eeqar
 The total disk luminosity is
 \beqar
 L_{disk} & = & \int_{0}^{\infty}2 \pi r F(r) dr = L_{t} 
 \frac{4+3\eta}{8} \\
    & \simeq & 0.8086 L_{t}. \nonumber \label{ldisk}
 \eeqar
 It represents thus the main part of the total bolometric luminosity 
 $L_{t}$.

 \subsection{Disk emission spectrum}
 Under the assumption of isotropic blackbody emission, without limb darkening,
  the emitted spectrum, integrated over angles, can be written as
 \begin{equation}
        L_{\nu} =  \int_{r_{min}}^{r_{max}} \pi B_{\nu}(T_{eff}) 
2 \pi r dr
 \end{equation}
 where the effective temperature $T_{eff}$ is defined in equation 
 (\ref{emiss}) and $ \displaystyle B_{\nu}(T) = \frac{2 h \nu^3}{c^2}
 \left[exp\left(\frac{h\nu}{k_B T_{eff}(r)}\right)-1\right]^{-1}$ is the
 usual Planck source  
function.
\noindent
Introducing the characteristic variables: 
\beqar
T_c & \equiv & \left(\frac{3 L_t}{32 \pi \sigma Z_0^2}\right)^{1/4} 
\label{Tc} \\
\nu_c & \equiv & k_B T_c/h \label{nuc}
\eeqar
and defining the following reduced variables:
\beqar
\overline{T} & = & T/T_c    \nonumber\\
\overline{\nu} & = &\nu/\nu_c \label{reduced} \\
\overline{L}_{\overline{\nu}} & = & \frac{\nu_{c}}{L_{t}} L_{\nu}, 
\nonumber
\eeqar
one can write the emitted spectrum in the universal form:
 \beqar
    \overline{T}(\mu') & = & \mu^{'3/4} F_{\eta\chi}(-\mu')^{1/4} \\
    \overline{L}_{\overline{\nu}}&= &  \frac{45 \overline{\nu}^3}{16
    \pi^{4}}\int_{\mu'_{min}}^ 
    {\mu'_{max}}\mu^{'-3}\left[\exp\left( 
        \frac{\overline{\nu}}{\overline{T}(\mu')}\right)-1\right]^{-1}d\mu'.
 \eeqar

 It is interesting to compare this spectrum with that of a standard SS
 accretion disk. 
  The flux emitted by a SS 
disk with a null torque at the inner radius $r_{i}$ is given by
\beq
 F(r) = \sigma T^{4}_{eff,SS}(r) = \frac{3 r_{i} L_{disk}^{(SS)}}{2 \pi r^{3}} 
 \left(1-\left(\frac{r_{i}}{r}\right)^{1/2}\right). \label{specSS}
\eeq
For large $\nu$, emitted at large distances, the spectra have the same 
 slope, with the dependence 
 $L_{\nu}\propto \nu^{1/3}$. This is because the release of gravitational 
 energy and the illumination by a central source give rise to the same 
 dependence of the energy flux at large distance $F(r) \propto 
 r^{-3}$.\\
Inspection of formulae (\ref{flux}), (\ref{ldisk}) and (\ref{specSS})
 show 
that the low frequency luminosity will be the same for both models if
\beq
 \frac{L_{disk}}{L_{disk}^{(SS)}} = \frac{2(4+3\eta)}{(3-\chi)} 
\left(\frac{r_{i}}{z_{0}}\right) \label{rapSS}
\eeq
For a Schwarzschild black hole with a mass $M$, $ r_{i} = 6 GM/c^2$. 
 \begin{figure}
 \psfig{width=\columnwidth,file=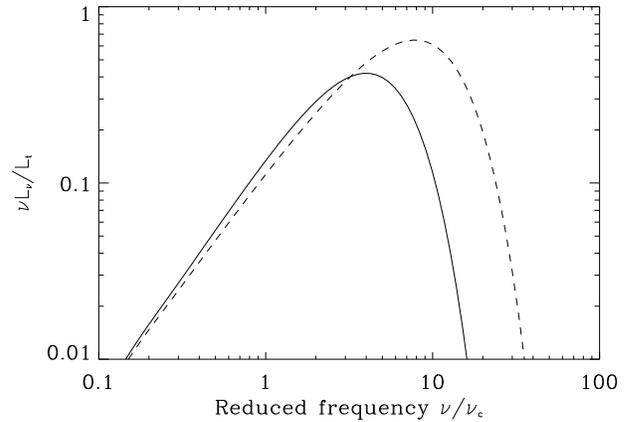 }
\caption[]{Spectra of the present illumination model for an 
        infinite slab (plain line) and a standard SS accretion disk 
        in reduced units (see text for definitions). The 2 spectra are
        normalized in order to have the same low frequency flux}\label{spec}  
 \end{figure}

 \begin{figure}
 \psfig{width=\columnwidth,file=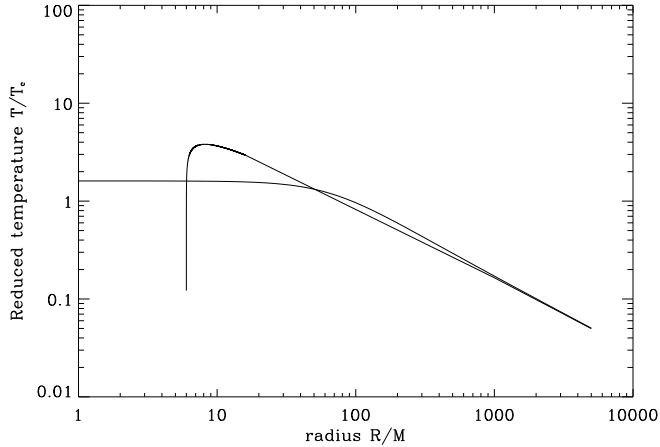 }
\caption[]{Radial temperature profile of the present illumination model for an 
        infinite slab (plain line) and a standard SS accretion disk
        (dashed line). The 2 curves are also normalized in order to have
        the same asymptotic behaviors for $r\rightarrow\infty$ 
           }\label{tdisk}  
 \end{figure}
\noindent
For comparison, we have chosen $ Z_{0} = 70 GM/c^{2}$.  
Figure (\ref{spec}) represents the exact spectrum emitted by the disk in
our illumination model, with 
$\mu'_{min}=0$ and $\mu'_{max}=1$, and in the standard SS model.  Figure (\ref{tdisk}) represents  
 the corresponding radial temperature profile for both models. As one can
 see, in the illumination model, the 
 temperature stops rising at a distance $r \sim Z_0$ whereas it keeps 
 growing in the standard accretion disk model, where most of accretion energy 
 is released at the smallest radii. As a consequence, for the same low 
 frequency flux, the illumination model predicts a lower bolometric 
 luminosity, and peaks at a lower frequency than the standard SS model.

 \subsection{Ratio of high energy to disk emission}
 The model predicts also a definite ratio between the high energy IC 
 luminosity and the disk thermal emission; the total (non intercepted) 
 IC luminosity is emitted in the upper hemisphere ($ 0 \leq \mu \leq 1$), 
 whereas the disk luminosity is equal to the IC luminosity in the lower 
 hemisphere ($ -1 \leq \mu \leq 0$). From equation (\ref{dpdo2}), one gets
 \begin{eqnarray}
        R & = &\frac{ L_{he}}{L_{disk}} = 
\frac{\int_{0}^{1}\frac{dP}{d\Omega} 2 \pi d\mu}
      {\int_{-1}^{0}\frac{dP}{d\Omega} 2 \pi d\mu} \nonumber \\
      & = & \frac{1- 3 \eta/4}{1 + 3 \eta/4}\label{ratioR}
 \end{eqnarray}
  One has $R \sim 0.237$ for an infinite plane. One can also evaluate the ratio 
  between the apparent disk and high
  energy luminosities in a given direction of observation $\mu_0$. It is given by:
 \beqar
        R(\mu_0) & = & \frac{\pi (dP/d\Omega)_{IC}}{\mu_0 L_{disk}}
\nonumber \\
        & = & \frac{3F_{\eta\chi}(\mu_0)}{4\mu_0(4+3\eta)}.
 \eeqar 
 \begin{figure}
  \psfig{width=\columnwidth,file= 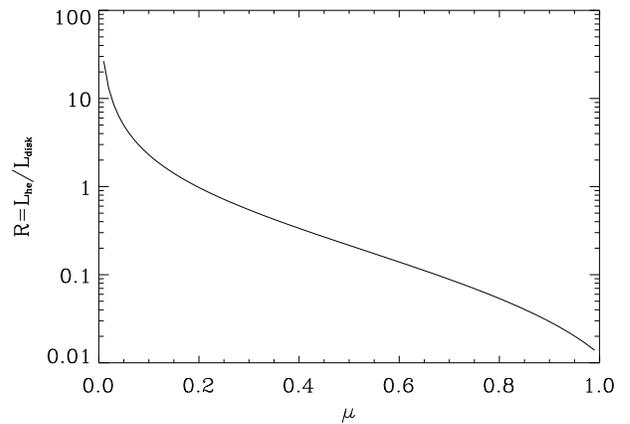}
\caption[]{Ratio of high energy to soft energy luminosity as a function
  of inclination angle, in logarithmic scale. The disk is modeled by an
  infinite slab}\label{ratio}  
 \end{figure}

This function is displayed in Fig. (\ref{ratio}). It presents a 
 minimum on the disk axis ($\mu_0 = 1$), both because the disk emission 
 is maximal at small inclination angle, and IC is minimal because of 
 the soft photon anisotropy. Thus, the model predicts that the 
{\it observed}
 X/UV ratio is smaller than 1 for $\mu_0 \geq 0.2$, and can be as low as 
 0.012. 

 \section{Application to AGN}
\label{AGN}
 \subsection{The high energy spectrum}
 As reminded in the introduction, the high energy emission of Seyfert 
 galaxies can be well reproduced by a primary power-law spectrum with a 
 spectral index $\alpha \simeq 0.9$,
 superimposed on more complex structures, that can be produced by the 
 reflection of the hard X-rays on a cold surface. Although the precise 
 modeling of such a reflection component is beyond the scope of this 
 paper, it is clearly compatible with the above picture. The primary 
 power-law emission should be associated with the emission from the hot 
 source, and the UV-optical component (Blue Bump) associated with the 
 reprocessed radiation, together with a Compton backscattered component 
 in X-rays (not taken into account in the present model). Noticeably, 
this power law is exponentially cut-off above a 
 characteristic energy of about 100 keV, with some uncertainty its
 precise value. Contrarily to thermal models, 
 where this cut-off is related to the temperature of the hot comptonizing 
 plasma, we propose to interpret it as a high energy cut-off of the 
relativistic 
 energy distribution. If the UV bump is located around 10 eV, this gives 
 an cut-off Lorentz factor $\gamma_0 \sim 10^2$. Although a detailed 
 model of the high energy source is again out of the scope of this work, 
 one can note that a model associating pair production and pair 
 reacceleration can provide such upper cut-off, to avoid catastrophic 
 run-away pair production (Done et al. \cite{Done90} ; Henri \& 
 Pelletier \cite{HP91} ).\\
 To account for the high energy cut-off, we will assume that the 
 particles (electrons or positrons) energy distribution function
 (integrated over volume) has the form:
 \beq
    n(\gamma)  =  N_{0} \gamma^{-s} 
    \exp\left(-\frac{\gamma}{\gamma_{0}}\right),\  \gamma_{min} <
    \gamma < \gamma_{max} \label{energydist}
 \eeq
 Inserting this function in Eq. (\ref{spectrehe}), noting that
 $f(\gamma,\mu) = n(\gamma)/2$, one gets the 
 expression of high energy specific power:
 \beqar
 \frac{dP}{d\Omega d\nu} & = & \frac{N_{0}  \sigma_{T}}{4} J 
\feta \left(\frac{\langle \epsilon_s \rangle 
 A(\mu)}{h}\right)^{(s-3)/2} \nu^{-(s-1)/2} \times \nonumber \\ 
 & & \exp\left(-\left(\frac{\nu}{A(\mu)\nu_0}\right)^{1/2}\right) 
 \label{dPdodnu}
  \eeqar 
 where $ \nu_0 = \gamma_{0}^{2}\langle \epsilon_{s}
  \rangle/h$ is a high energy frequency 
  cut-off.\\ 
  Integrating Eq. (\ref{dpdo}) over all angles, one can also write 
  the total high energy luminosity as:
  \beq
      L_{t} = \frac{16\pi}{3}C_{0} N_{0}\sigma_{T}J \gamma_0^{(3-s)} 
  \eeq
  where $C_0$ is expressed as an incomplete gamma function: 
  \beq
    C_{0} =
    \int_{\gamma_{min}/\gamma_{0}}^{\gamma_{max}/\gamma_{0}}x^{2-s}\exp(-x)dx = \Gamma(3-s;\gamma_{min}/\gamma_0; \gamma_{max}/gamma_0)
  \eeq
  One can also evaluate the mean photon energy
  \beq
  \langle \epsilon_{s} \rangle = \frac{J}{\oJ}= C_{1} h \nu_{c} 
  \label{meanenergy}
  \eeq
  where
  \beq
   C_{1} = \frac{8 \zeta(4)}{3\xi \zeta(3)\int_{\mu_{min}}^{\mu_{max}}
     \oT^{3}d\mu}     
  \eeq
  where $\zeta(n)$ is the Riemann Zeta function.
  One can finally compute the photon number anisotropy parameter $\oeta$,
  appearing in the computation of $A(\mu)$ (cf. Eq. (\ref{amu})):
  \beq
  \oeta = \frac{\oH}{\oJ} = {\frac{\int_{\mu_{min}}^{\mu_{max}}\mu
      \oT^{3/4}d\mu}{\int_{\mu_{min}}^{\mu_{max}} \oT^{3/4}d\mu}}.
  \eeq
  remembering that the specific intensity emitted by the disk is supposed
  to be the Planck function.
  For $\mu_{min} = 0$ and $\mu_{max} = 1$, one gets $C_{1} \simeq 
  3.41$ and $\oeta \simeq 0.788$
  Using the reduced variables defined by Eq. (\ref{reduced}), one 
  gets finally the following expression for the high energy specific power:
  \begin{eqnarray}
  \frac{d \overline{P}}{d\Omega d\overline{\nu}} &=& 
   \frac{3\gamma_0^{s-3}}{64\pi C_{0}} 
   \left(C_1A(\mu)\right)^{(s-3)/2} 
   \overline{\nu}^{(1-s)/2}\times\\ \nonumber
                                                 & &
   \feta \exp\left[-\left(\frac{\overline{\nu}}{A(\mu)\overline{\nu}_0}
   \right)^{1/2}\right]\label{Lreduced}
  \end{eqnarray}
   with $ \overline{\nu}_0 = \nu_{0}/\nu_{c}$.
   Noticeably, the high energy cut-off depends on the inclination 
   angle, the smallest angles giving the lowest cut-off.
  
\begin{figure}
  \psfig{width=\columnwidth,file=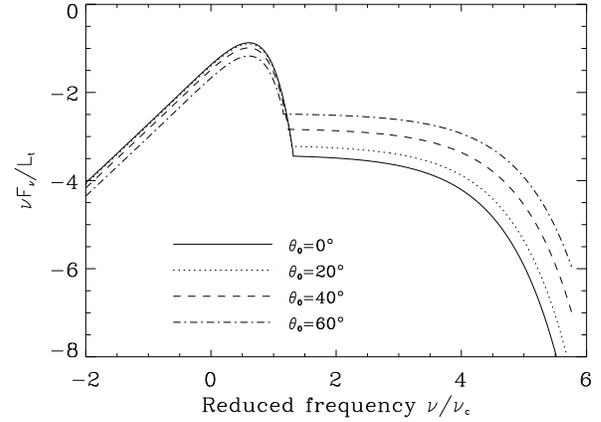 }
\caption[]{Broad band spectra for various inclination angles, in reduced
  units. The values of parameters are the following: $s =3$,
  $\overline{\nu}_0 = 10^4$}\label{specang}  
 \end{figure}
 
 \subsection{Validity of the point source approximation}
It is obvious that a point source 
is a convenient, but unrealistic approximation of the real geometry of 
the source, since it has a zero cross section and a infinite Thomson 
opacity. Rather, for a given number $N$ of scattering particles, 
one gets a 
minimal size $R_{min}$ for the source being optically thin; for a homogeneous 
sphere, it requires
\beq 
R \geq R_{min} = \left(\frac{3N\sigma_{T}}{4\pi}\right)^{1/2} \label{rmin}
\eeq

This hot  
source will sustain a solid angle $\Omega \simeq \pi (R_{min}/Z_0)^{2}$. 
The total number of particles in the optically thin regime can be 
calculated by Eq. (\ref{ksi}) and (\ref{Lt}), which yields:
\beq
N = \frac{2 \pi Z_{0}^{2} \xi}{\sigma_{T}\langle\gamma^2\rangle} 
\label{npart} 
\eeq
Using the distribution function given by Eq. (\ref{energydist}), one 
gets
\beq
\langle\gamma^2\rangle = 
\gamma_0^{2}\frac{\Gamma(3-s;\gamma_{min}/\gamma_0;\gamma_{max}/\gamma_0)
}{\Gamma(1-s;\gamma_{min}/\gamma_0;\gamma_{max}/\gamma_0)} \label{gamma2}
\eeq

Combining Eq. (\ref{rmin}), (\ref{npart}) and (\ref{gamma2}), one obtains a 
estimate of the minimal radius of the source:

\begin{equation}
        \frac{R_{min}}{Z_{0}} = \frac{1}{\gamma_{0}} \left(\frac{3 \xi
 \Gamma(1-s;\gamma_{min}/\gamma_0;\gamma_{max}/\gamma_0)
}{2\Gamma(3-s;\gamma_{min}/\gamma_0;\gamma_{max}/\gamma_0)}\right)^{1/2}
        \label{anglemin}
\end{equation}
 \begin{figure}
\psfig{width=\columnwidth,file=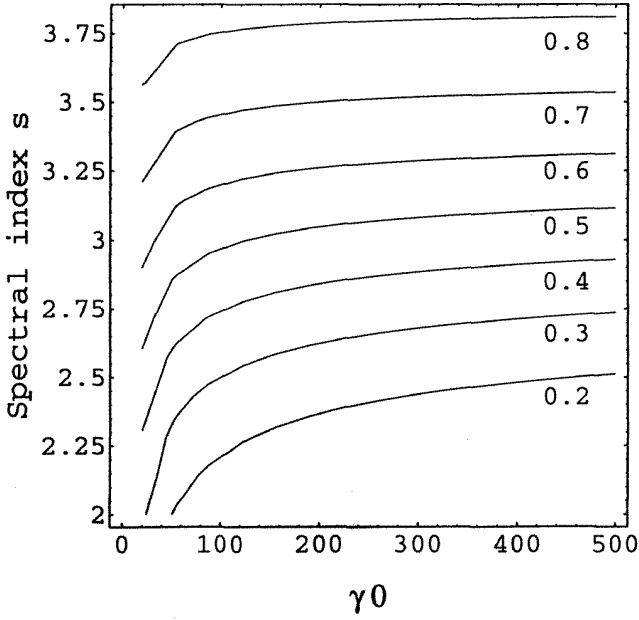 }
\caption[]{Contour plot of the minimal angular size of the source 
$R_{min}/Z_{0}$, as a function of the spectral index $s$ and the 
cut-off Lorentz factor $\gamma_{0}$. The limits of the particle 
energy distribution are $\gamma_{min} = 1$ and $\gamma_{max} = 
\infty $}  \label{rminfig}  
 \end{figure}
Figure (\ref{rminfig}) represents the contour plot of $R_{min}/Z_{0}$ as a 
function of $\gamma_{0}$ and $s$, for $\gamma_{min} = 1$ and $\gamma_{max} = 
\infty$. As we have mentioned, high energy observations of Seyferts 
seem to favour values of $\gamma_{0} \simeq 10^{2}$, and $s \simeq 3$. 
This corresponds to $R/Z_{0} \simeq 0.5$, which means that the hot 
source is not really point-like, but moderately extended: it could be 
realized by a region with a radius $R \sim 15 r_{g}$, located around 
$Z_{0} \sim 30 r_{g}$. These values seem reasonable for a shock in 
the inner region of a jet emitted by an accretion disk. Of course, a 
correct treatment should take into account the finite size of the 
source, but the calculations are much more involved if the particles 
are off-axis: the present theory must be considered as a preliminary 
one, and the application to the extended case will be treated in a 
future work.
 
 \subsection{Application to broad-band spectra.}
 Figure (\ref{specang}) represents the overall spectra predicted by the 
model for different inclination angles, ranging from $\theta_0 = 0^\circ$ 
to $60^\circ$. For all these angles, the X-ray luminosity is apparently 
 smaller than the UV bump. Higher inclination angles will probably lead
 to strong absorption through the external parts of the disk, presumably a 
 molecular torus: in the unification scheme, they would correspond to 
 Seyfert 2 galaxies (but see below).
 The UV/X ratio depends markedly on the inclination
 angle: noticeably, the smaller ratio correspond to almost face-on 
objects. The upper energy cut-off is also lightly angle 
 dependent, due to the $ A(\mu)$ factor in Eq. (\ref{Lreduced}). 
 Face-on objects have the lowest cut-off.\\
 Direct comparison with
 observations is difficult at this stage, because the model does 
 not include other important components, like the absorption edge, 
the Compton reflection feature 
 and the fluorescent Fe K$\alpha$ line that are observed in many 
 Seyferts. 
Without detailed calculations, it can be expected however that
 these components are more pronounced that those predicted by an
 isotropic illumination model, the mean enhancement factor being 
of the order
 of $R^{-1} \simeq 4$ (Eq. (\ref{ratioR})). Ginga and more recent
 ASCA observations (Nandra et al. \cite{Nand97}) have found large
 equivalent widths for the Fe K$\alpha$ line, with a mean value around 
$230 \pm 60$ eV, but up to $550^{90}_{-120}$ eV for NGC 4151, 
whereas detailed calculations of George and Fabian
(\cite{GF91}) and Matt et al. (\cite{MattP91}) predicted a value around 140
eV. These results depend however on the ionization state of the
matter (Matt et al. \cite{MattF93}) and the assumed iron abondance
(Reynolds et al. \cite{Reyn95}). Clearly more work must be done to
clarify this issue.\\
The resolved line profiles seen by ASCA can be well fitted
by an illuminated accretion disk in rotation around a black hole, with
$\simeq 50 $ per cent of the line emission originating within $\simeq 20$
$r_g$, and $\simeq 80 $ per cent within $\simeq 100$ $r_g$(Nandra et al.
 \cite{Nand97}): this is thus
compatible with a hot source located at a few ten $r_g$ above the 
disk.\\
Although a precise prediction of UV/X flux is difficult to assess, we have 
used the work of Walter \& Fink (\cite{WalF93}), who compared the X 
and UV flux in a large sample of Seyfert 1 galaxies, to make a rough 
comparison with our model. We assumed that the $1375$ $\AA$ flux is 
close to the maximum of the UV bump, whereas the 2 keV flux is not 
strongly contaminated by the soft excess or the reflection component, 
and is representative of the low energy part of the hard X-ray 
spectrum given by Eq. (\ref{dPdodnu}). We thus consider the apparent UV/X
ratio $R_a$ of these two values.\\
In our model, the maximum of the UV bump is found numerically to be 
$ \overline{\nu}\overline{L}_{\overline{\nu},max} \simeq 0.419$ at
${\overline{\nu}} \simeq 4$. 
With the geometrical factor due to disk inclination, one gets thus

\beq
  \overline{\nu}\frac{d\overline{P}_{UV}}{d\Omega d \overline{\nu}}(\mu)
  \simeq \frac{0.419}{\pi} \mu
\eeq
 
The hard 
X-ray flux at low energy is given by Eq. (\ref{dPdodnu}) with
 $\overline{\nu} \ll {\overline{\nu_{0}}}$ Taking $s=3$ for simplicity, 
 one gets
 \beq
  \overline{\nu}\frac{d \overline{P}_X}{d\Omega d\overline{\nu}} (\mu) \simeq 
  \frac{3\feta}{64 \pi C_{0}}
  \eeq
  , so that:
  \beq
    R_{a}(\mu) \simeq 9.05 \frac{\mu C_{0}}{\feta} 
  \eeq
  Now for a sample of $N_{g}$ galaxies oriented randomly between $\mu =0$
  and $\mu =1$, 
  one expects a uniform distribution
  \beq
 \frac{dN}{d\mu} = N_{g}
  \eeq

One gets the probability distribution of the UV/X ratio $R_{a}$:

  \beq
\frac{dP}{dR_a} =  \left|\frac{d\mu}{dR_a}\right|=
\frac{\feta^2}{9.05 C_0(3-\chi-(3\chi-1)\mu^2)} 
  \eeq
\begin{figure}
  \psfig{width=\columnwidth,file=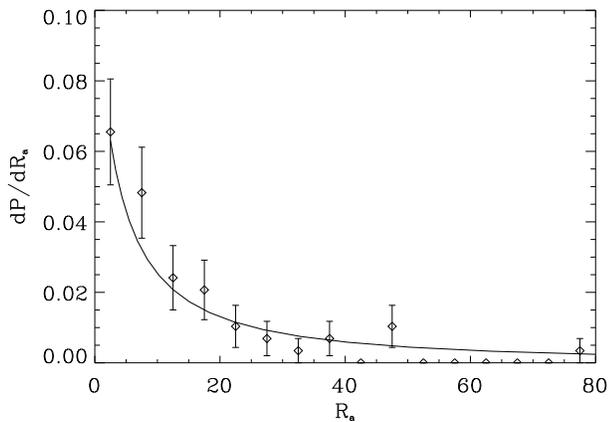 }
\caption[]{Comparison of the probability distribution of
apparent UV/X ratio $R_a$ compiled by Walter
  and Fink(\cite{WalF93}) and the predicted values, with $\gamma_0 = 10^2$}
\label{historatio}  
 \end{figure}
  
  Figure (\ref{historatio}) represents the comparison of observed and 
  predicted probability distribution of the apparent 
UV/X ratio $R_{a}$, for 
$\gamma_{0} = 10^{2}$. To derive the probability density $dP/dR_a$,
 the observed values have been binned into 
  intervals of width 5, and statistical error bars have been 
  added. The agreement is very satisfactory, which can be rather
  fortuitous in view of the approximations used.\\
Although the widespread opinion, based on unification models, is that
Seyfert 1 galaxies are seen nearly pole-on, whereas Seyfert 2 are their
edge-on counterparts, there is some evidence that the reality may be more
complex. In a statistical study of the 48 Seyfert galaxies from the CfA
catalog, Edelson et al. (\cite{Edel87}) concluded that all Seyfert 2, but
only one third of Seyfert 1,
present an excess at 60 $\mu m$, attributed to thermal emission from
dust. Since emission from dust is not expected to be highly anisotropic,
it would imply that the unification model applies only to a subclass of
objects, which possess an obscuring torus with some opening angle. The
other subclass could not possess this torus, and give only Seyfert 1
galaxies seen under any inclination angle. This could explain the presence
of high inclination angles, i.e. low UV/X ratios galaxies. But of course,
a large intrinsic or extrinsic absorption could change a lot this picture.
 \subsection{Scaling laws} 
 \label{Discussion}
 As is well known, a usual assumption for the mass-luminosity ratio 
  of AGN is that the bolometric luminosity is limited by the radiation 
  pressure to the Eddington luminosity:
  \beq
  L_{E} = \frac{4\pi G M m_{p} c}{\sigma_{T}} \label{ledd}.
  \eeq
This relationship predicts a linear correlation between mass and 
luminosity for Eddington accreting black-holes, $L \propto M$; 
the corresponding Eddington temperature is 
given by 
\beq
T_{E} = \left(\frac{L_t}{4\pi r_g^2}\right)^{1/4} \propto M^{-1/4}. 
\eeq
We show here that the present model predicts different scaling laws, if
one adds some  
supplementary assumptions on the physics of the hot source. At first 
sight, all equations of the models are linear with respect to the 
global luminosity, so no particular relationship is predicted between 
the luminosity and other parameters like $M$ or $Z_0$. However things are 
different if one considers the microphysics and a realistic 
geometry of the hot source. First, let us consider again the upper 
energy cut-off of the IC spectrum discussed in Paragraph (\ref{paraappli}). 
Observations seem to show that all Seyfert galaxies have the cut-off 
around the same value, approximately 100 keV; however, when taking 
into account the reflection component, this estimate could be somewhat 
higher, 
up to 400 keV (Zdziarski et al. \cite{Zdz95}). This is sufficiently close 
to pair production threshold 
to make plausible the idea that this cut-off is in some way fixed by a 
regulation mechanism to avoid run-away pair production. In this case, 
the cut-off energy is a physical quantity, determined by microphysics 
rather that macrophysical quantities. The maximal energy of photons 
produced by IC process is of the order
\beq
\gamma_{0}^{2} h \nu_{c} \approx constant. \label{gamma0cst}
\eeq

 Now it is plausible that the size of the source and its distance to 
 the black hole are controlled by the global environment responsible 
 for the hot source. Conceivably, it could be realized through a 
 strong shock terminating an aborted jet. If one makes the (admittedly 
 crude) assumption that all distances scale like the hole radius 
 $r_{g}$, then one gets $ R/Z_{0} \approx constant$. For a given 
 spectral index $s$, Eq. (\ref{anglemin}) predicts $ \gamma_{0} \approx constant$,
  and Eq. (\ref{gamma0cst}) gives $\nu_{c} \approx constant$. Finally one 
 gets the following scaling laws:
 \begin{eqnarray}
         T_{c} & \propto & \nu_c \propto M^{0} \approx constant  \label{Tm0} \\
         L_{t} & \propto & M^{2}. \label{Lm2}
 \end{eqnarray}
 that is , the temperature of the Blue Bump does not depend on the 
 mass and the luminosity scales like the square of the mass.
 Of course, one could observe substantial variations of at least one 
 of these quantities if $s$ or $\Omega$ varies, either by a variation of 
 $Z_0$ or a variation of $R$.
 Interestingly, observations seem to corroborate these behaviors: on 
 a sample of many quasars and galaxies spanning a large range of 
 masses, Walter \& Fink (\cite{WalF93}) found that the Blue Bump 
and soft X-ray 
 excess were approximately in the same ratio, 
although the central masses can differ 
 by a factor $10^{4}$. This is very hard to explain in the frame of 
 conventional accretion models. In another study, using a specific model 
 to describe the width of Broad Lines from the emission by the 
external part of the disk, Collin-Souffrin \& Joly  (\cite{ColJol91}) 
have 
 deduced the inclination angles and central masses of a sample of 
 Seyfert 1 galaxies and quasars. They found a correlation between mass 
 and luminosity under the form $ L \propto M^{\beta}$, with $\beta = 1.8
 \pm 0.6 $.
 Although these results have been obtained on a limited sample, they 
 are compatible with the previous results and clearly 
 different from those predicted by the standard accretion disk models. 
 A rather intriguing consequence is that there is a maximal mass above 
 which the accretion becomes impossible by such a mechanism, 
where the luminosity 
 predicted by Eq. (\ref{Lm2}) gets higher than the Eddington 
 luminosity. From Eq. (\ref{Tc}) and (\ref{meanenergy}), the total luminosity
 can be written under the form:
 \beq
 L_{t } = \frac{256 \pi^{6}G^{2}}{45 C_{1}^{4}h^{3}c^{6}}
  \left(Z_{0}/r_{g}\right)^{2} M^{2} \langle \epsilon_{s} \rangle^{4} 
 \eeq
 Comparing with Eq. (\ref{ledd}), one finds that $L_{t} = L_{edd}$ for
 
\beqar
M & = & \frac{45 C_{1}^{4}h^{3}c^{7}m_{p}} {64 \pi^{5}G \sigma_{T}}
  \left(Z_{0}/r_{g}\right)^{-2} \langle \epsilon_{s} \rangle^{-4} \\
  & = & 10^{10} \left(\frac{Z_{0}}{ 30 r_{g}}\right)^{-2} 
 \left( \frac{\langle \epsilon_{s} \rangle }{ 5 \rm{eV}} 
 \right)^{-4} M_{\sun} \label{Mmax}
 \eeqar
 
 This value is surprisingly close to that commonly invoked for the 
 most luminous known quasars, which apparently accrete at a 
 near-Eddington rate. It is however unclear how accretion would be 
 stopped for higher central masses, since the radiation pressure can 
 be effective only if the central engine actually radiates.

\section{ Conclusion } 
   We have shown that a model based on reillumination of a disk by an 
   anisotropic IC source could lead to a self-consistent picture 
   where the angular distribution of high energy radiation and the 
   radial temperature distribution of the disk are mutually linked and 
   both  
   determined in a single way. The model offers a simple explanation 
   for the correlated long term variability of X and UV radiation, 
   the short term variability of X-rays non correlated with UV 
   variations, and the apparent X/UV deficit that seems 
   contradictory with simple reillumination models. In its simplest 
   form, it predicts a unique shape of disk spectrum and a X/UV ratio 
   depending only on the inclination angle. The predicted values are in 
   good agreement with observations. In an accompanying
   paper (Petrucci \& Henri \cite{PH97}), we shall study the influence of
    relativistic corrections on the 
   above scheme to account for the gravitational effect of a 
   Schwarzschild or a Kerr black hole. A precise comparison with real 
   spectra should also include other components, such as a 
   reflection component and a fluorescent Fe K$\alpha$ line. This is 
   deferred to a future work.


\end{document}